\documentclass[fleqn,10pt]{wlscirep}
\usepackage[utf8]{inputenc}
\usepackage[T1]{fontenc}
\usepackage{amsmath,amssymb,amsthm,graphicx,braket,multirow,authblk,amsthm,dcolumn,latexsym,subcaption,mathtools,cite,cancel}
\usepackage[normalem]{ulem}
\DeclareSymbolFontAlphabet{\amsmathbb}{AMSb}
\usepackage{graphicx}
\usepackage{amsmath,amssymb}
\usepackage{mathtools,xcolor}
\usepackage{float}
\usepackage{authblk}
\usepackage{graphicx}
\usepackage{amsmath,amssymb}
\usepackage{mathtools,xcolor}
\usepackage{float}
\usepackage{authblk}
\usepackage{bm}
\newcommand{\tr}{\operatorname{Tr}}

\usepackage{cite}
\usepackage[nottoc]{tocbibind}
\newcommand{\la}{\langle}
\newcommand{\ra}{\rangle}
\def\<{\langle}
\def\>{\rangle}

\newcommand{\be}{\begin{equation}}
\newcommand{\ee}{\end{equation}}
\newcommand{\ba}{\begin{eqnarray}}
\newcommand{\ea}{\end{eqnarray}}

\def\h{\hskip 1cm}

\def\lo{\longrightarrow}
\def\a{\alpha}

\def\ni{\noindent}

\newtheorem{proposition}{Proposition}
\newtheorem{COR}{Corollary}
\newtheorem{EX}{Example}

\newtheorem{theorem}{Theorem}

\title{A class of entanglement witnesses and a realignment-like criterion}

\author[1]{Vahid Jannesary}
\author[1]{Vahid~Karimipour}
\author[2]{Dariusz~Chru\'sci\'nski}
\affil[1]{Department of Physics, Sharif University of Technology, Tehran 14588, Iran}
\affil[2]{Institute of Physics, Faculty of Physics, Astronomy and Informatics, Nicolaus Copernicus University, Grudziadzka 5/7, 87-100 Torun, Poland}

\begin{abstract}

We present a  multi-parameter family of positive maps between spaces of differing dimensions. This framework facilitates the construction of Entanglement Witnesses (EWs) specifically designed for systems living in $d_1\times d_2$
dimensions. A sufficient condition for positivity is presented. Interestingly, it is shown that all EWs constructed this way are equivalent to a single realignment-like criterion which for $d_1 \neq d_2$ is in general stronger than the original realignment criterion. We illustrate effectiveness of this criterion considering examples of Positive Partial Transpose (PPT) entangled states in $3 \times 3$ and $3\times 4$ dimensions.

\end{abstract}

\begin{document}

\flushbottom
\maketitle
%
%
\thispagestyle{empty}



	





\date{}


\section{Introduction}

Quantum entanglement is one of the crucial resource for modern quantum
technologies like quantum cryptography, quantum communication, 
and quantum computation\cite{QIT,HHHH}. One of the main problem of the
entanglement theory is to devise appropriate criteria which
enables one to distinguish separable and entangled states \cite{HHHH,Guhne}. The identification of entangled states, in particular, presents a formidable challenge, underscored by its classification as an NP-Hard problem in mathematical terms \cite{Gurvits2004}. Hence, the pursuit of viable techniques for detecting entangled states takes on paramount importance.
Entangled states are referred to as states that are not separable, and separable states refer to states that can be written as follows\cite{Werner1989}:
\begin{equation}\label{def1}
\rho = \sum p_i \rho_i^{(1)} \otimes \rho_i^{(2)} ,
\end{equation}
where $0 \leq p_i \leq 1$ and $\sum p_i = 1$.
It is not easy at all to determine whether or not a given state  can be expressed in the aforementioned form.  As a result, there is a pressing need for more effective methods for the detection of entanglement. Over time, diverse techniques have emerged for this purpose.   For low-dimensional bipartite
systems, i.e. qubit-qubit and qubit-qutrit, the separability poblem is completely solved by the well known Peres-
Horodecki criterium which states that a state is separable if and only if it is
positive under partial transpose (PPT), that is, $\rho^\Gamma = (I \otimes T)\rho \geq 0$ \cite{Peres1996,Horodecki1996}. However, for higher dimensional systems  the separability problem is notoriously
difficult. In higher dimensions one should search for other criteria of entanglement, many of which have been reviewed in \cite{Guehne2009}. The most important ones are called Entanglement Witnesses (EW), (cf. \cite{Chruscinski2014} for a review). An EW refers to a Hermitian operator  $W$, which has negative eigenvalues but has the following property:
\begin{equation}\nonumber
	\tr(W \sigma) \geq 0, \quad \forall \sigma \in S,
\end{equation}
where $S$ is the set of all separable states. Therefore, a negative expectation value for $W$ on  a state $\rho$, then means that $\rho$ is entangled. Alternatively, one can think of $W$ as a block-positive operator, i.e.
\be\nonumber
\langle \psi, \phi | W | \psi, \phi \rangle \geq 0 ,\ \forall \ \psi \in H_1 \ , \forall \ \phi \in H_2,
\ee
which  has negative eigenvalue when acting on the whole Hilbert space $H_1\otimes H_2$.
Hence, any entangled state $\rho$ of a bipartite system can be detected a suitable entanglement
witness such that $Tr(W\rho) < 0$  \cite{HHHH,Terhal2002,Chruscinski2014,Guehne2009}. 

The well-known criterion based
on positive maps \cite{HHHH,Horodecki1996} states that if $\rho$ is separable, then $(I \otimes \Phi)\rho \geq 0$ for all positive maps $\Phi$. Actually, it is well known that any entanglement witness $W$ can be written in the form
\begin{equation}\nonumber
	W = (I \otimes \Phi) |\phi^+ \rangle \langle \phi^+| ,
\end{equation}
where $\Phi$ is a positive (but not completely positive) map, and $|\phi^+\rangle = \frac{1}{\sqrt{d}}\sum_{\mu=0}^{d-1} |\mu \mu \rangle$,  is a maximally entangled state\cite{Horodecki1996}. In other words, $W$ is the Choi-matrix of the map $\Phi$, which due to its lack of complete positivity can have negative eigenvalue.  

Apart from positive maps or entanglement witnesses criteria there exist other well known effective methods for entanglement detection. One example is a realignment (or computable crossnorm
(CCNR)) criterion \cite{Chen1,Chen2,Rudolph}. This criterion proved to be very effective in the detection of PPT entangled states \cite{HHHH} (cf. the recent review \cite{Beatrix1}).

In this paper we  construct a large class of positive maps which satisfy simple condition which does guarantee the positivity of the map. Equivalently, we provide a large class of entanglement witnesses in $d_1 \otimes d_2$ quantum systems. The structure of this paper is as follows: In Section \ref{RESULTS} we present the main results (Theorem 1). Interestingly, several well known maps/witnesses belong to this class. We illustrate this result by several examples. Moreover, we relate our result to so called mirrored witnesses \cite{Bae2020,Anindita2023}. Section \ref{REAL} shows that our class of witnesses is essentially equivalent to a single realignment-like criterion (Theorem 3). The power of this criterion is then illustrated by two examples of $4 \otimes 4$ and $3\otimes 4$ systems.  Final conclusions are collected in Section \ref{CON}.

\section{Results}  \label{RESULTS}
 \subsection{Notations and Conventions}\label{notn}
 
 Let $H_d$ be the Hilbert space of dimension $d$ and   $L(H_{d})$  the space of linear operators acting on $H_{d}$. The convex set of positive operators is denoted by $L(H_{d})^+$. A vector space of linear operators $L(H_{d})$
 is equipped with the Hilbert-Schmidt inner product $\la A, B\ra:=\tr(AB^\dagger)$.  
 In what follows we consider maps $\Phi: L^+(H_{d_1})\lo L^+(H_{d_2})$.
Consider two  orthonormal bases of Hermitian operators in  $L(H_{d_1})$:  

$$ \left\{\Gamma_0=I_{d_1}/{\sqrt{d_1}},\  \Gamma_i; \ i=1,\ldots, d_1^2-1 \right\} , \ \ \tr(\Gamma_\a\Gamma_{\a'})=\delta_{\a\a'} , $$
and  in $L(H_{d_2})$:

$$ \left\{\Omega_0=I_{d_2}/{\sqrt{d_2}},\  \Omega_k; \ k=1,\ldots,d_2^2-1\right\} , \ \ \tr(\Omega_\beta\Omega_{\beta'})=\delta_{\beta\beta'} , $$ 
where $\Gamma_i$ $(i>0)$ and $\Omega_k$ $(k>0)$ are traceless.
\ni
The state of a quantum system is denoted by $\rho\in D(H_{d})$, where $D(H_{d})\subset L(H_{d})^+$ is the convex subset of positive operators with unit trace.
Any Hermitian operator  $X \in L^+(H_{d_1})$  can be expanded as
\be\label{x0}
X= \sum_{\a=0}^{d_1^2-1} x_\a\Gamma_\a=\frac{x_0}{\sqrt{d_1}}I_{d_1}+{\bf x}\cdot \bm\Gamma,
\ee
where $x_0 \in \mathbb{R}$ and $ {\bf x}=(x_1,\ldots, x_{d_1^2-1})$ is a generalized Bloch vector. 
When $X$ is a rank-1 operator, it necessarily holds that $\tr X^2= (\tr X)^2$ which leads to
\be \label{ineq-x.x}
\| {\bf x}\|^2 = x_0^2 (d_1-1).
\ee
where $\| {\bf x}\|^2 =\sum_{i=1}^{d_1^2-1}x_ix_i$. Similarly, any Hermitian operator $Y$ in $L(H_{d_2})$ can be represented  as
\be\label{yy}
Y=  \sum_{\beta=0}^{d_2^2-1} y_\beta\Omega_\beta=\frac{y_0}{\sqrt{d_2}}I_{d_2}+{\bf y}\cdot \bm\Omega
.
\ee
\noindent
Consider a linear map $\Phi:L(H_{d_1})\to L(H_{d_2})$ defined by
\be\label{eq-orthonormal}
\Phi(\Gamma_\a)= \sum_{\beta=0}^{d_2^2-1} R_{\beta\alpha}\Omega_\beta,\ \
\ee
or equivalently
\be
\Phi(X) = \sum_\alpha \sum_\beta R_{\beta\alpha} \tr(\Gamma_\alpha X) \Omega_\beta ,
\ee 
where the matrix $R$ is parameterized as follows
\be
 R=\begin{pmatrix} R_{00} &{\bf s}\\ {\bf t}& \Lambda
\end{pmatrix}
\ee
with ${\bf t}\in R^{d_2^2-1} $,  ${\bf s}\in R^{d_1^2-1}$, and  $\Lambda\in M_{d_2^2-1\times d_1^2-1}$ is a real matrix. This map induces the following affine transformation on the level of generalized Bloch vectors

\be\label{affine}
\begin{pmatrix}y_0\\ {\bf y}\end{pmatrix}= \begin{pmatrix} R_{00} &{\bf s}\\ {\bf t}& \Lambda
\end{pmatrix}\begin{pmatrix}x_0\\ {\bf x}\end{pmatrix}.
\ee
Equivalently one has 

\be\label{Phix}
\Phi(X)=\Phi(x_0\Gamma_0+{\bf x}\cdot \bm\Gamma)=(R_{00}x_0+{\bf s}\cdot {\bf x})\Omega_0+(x_0{\bf t}+\Lambda {\bf x})\cdot \bm\Omega . \ee
In the special case one has:

\begin{enumerate}

\item $\Phi$ is  trace-preserving iff ${\bf s}=0$ and $R_{00} = \sqrt{\frac{d_1}{d_2}}$,

\item $\Phi$ is unital iff ${\bf t}=0$ and $R_{00} = \sqrt{\frac{d_2}{d_1}}$.

\end{enumerate}
Finally, let us recall the following result (known as a Mehta's lemma, cf. Refs. \cite{Mehta1989,mehta}).

\begin{proposition} \label{lem}
    Let $A$ be a Hermitian $d \times d$ matrix. If
	\be\label{lemma-mehta}
	\tr A^2 \leq \frac{(\tr A)^2}{d-1},
	\ee
	then $A$ is positive matrix.
\end{proposition}

\subsection{A class of positive maps} 
\label{sec-construction}

Given a quantum map $\Phi$ characterized by the affine transformation on the generalized Bloch vector ${\bf x}\lo \Lambda {\bf x}+x_0 {\bf t}$, our goal is to find the condition on this affine map which guarantees positivity of $\Phi$. 


\begin{theorem}
 A  map $\Phi$, implementing the affine transformation (\ref{affine}), is  positive if the following condition holds
 
\be\label{main_cond0}
\sqrt{d_2-1}\Vert {\bf t}\Vert+\sqrt{d_1-1}\Vert {\bf s}\Vert + \sqrt{(d_1-1)(d_2-1)}\Vert \Lambda\Vert_{\infty}
\leq R_{00} , 
\ee
where $\Vert \Lambda\Vert_{\infty}$ is the operator norm of $\Lambda$ and is equal to $\lambda_{max}$ which is the largest singular value of $\Lambda$.
\end{theorem}

\noindent
{\bf Proof:}  Due to linearity of the map, it suffices to prove the statement only for rank-1 positive operators $X$. For such $X$ (see \eqref{ineq-x.x}), we have $$\tr(\Phi(X))=y_0\sqrt{d_2}=(R_{00}x_0+{\bf s}\cdot {\bf x})\sqrt{d_2} $$ and
$$
\tr \Phi(X)^2 = \sum_{\beta} y_\beta y_\beta=y_0^2+{\bf y}\cdot {\bf y}=(R_{00}x_0+{\bf s}\cdot {\bf x})^2+\Vert x_0{\bf t}+\Lambda{\bf x}\Vert^2.
$$
Inserting these two expressions in Mehta's lemma we find that the sufficient condition for positivity of $\Phi(X)$ reads
\be
(R_{00}x_0+{\bf s}\cdot {\bf x})^2+\Vert x_0{\bf t}+\Lambda{\bf x}\Vert^2\leq \frac{(R_{00}x_0+{\bf s}\cdot {\bf x})^2d_2}{d_2-1}.
\ee
or, equivalently, 
\be\label{s}
\Vert x_0{\bf t}+\Lambda{\bf x}\Vert^2\leq \frac{(R_{00}x_0+{\bf s}\cdot {\bf x})^2}{d_2-1}.
\ee
Let us observe that
\be\label{s1}
\Vert  x_0{\bf t}+\Lambda{\bf x} \Vert\leq \Vert x_0{\bf t} \Vert + \Vert \Lambda {\bf x}\Vert\leq |x_0|\Vert {\bf t}\Vert +\Vert \Lambda\Vert_{\infty}\Vert {\bf x}\Vert=|x_0|\big(\Vert {\bf t}\Vert +\Vert \Lambda\Vert_{\infty}\sqrt{d_1-1}\big) ,
\ee
where in the last equality we have used the relation $\Vert {\bf x}\Vert=|x_0|\sqrt{d_1-1} $.
Moreover, 
\be
R_{00} |x_0|-\Vert {\bf s}\Vert \Vert {\bf x}\Vert \leq | R_{00} x_0+{\bf s}\cdot {\bf x}| ,
\ee
and hence  using again $\Vert {\bf x}\Vert=|x_0|\sqrt{d_1-1}$, one finds
\be\label{s2}
|x_0|\big(R_{00}  -\Vert {\bf s}\Vert \sqrt{d_1-1}\big) \leq | R_{00} x_0+{\bf s}\cdot {\bf x}| .
\ee
It is, therefore, clear that if 
\be
\Vert {\bf t}\Vert +\Vert \Lambda\Vert_{\infty}\sqrt{d_1-1} \leq \frac{ R_{00} -\Vert {\bf s}\Vert \sqrt{d_1-1}}{\sqrt{d_2-1}} ,
\ee
holds, then (\ref{s}) is trivially satisfied and hence the map $\Phi$ is positive. After rearranging terms one obtians 
\be
\sqrt{d_2-1}\Vert {\bf t}\Vert+\sqrt{d_1-1}\Vert {\bf s}\Vert + \sqrt{(d_1-1)(d_2-1)}\Vert \Lambda\Vert_{\infty}
\leq R_{00} 
\ee
which ends the proof.  \hfill $\Box$

\begin{COR}
 When $d_1=d_2=2$, this is simplified to
\be
\Vert {\bf t}\Vert+\Vert {\bf s}\Vert + \Vert \Lambda\Vert_{\infty}
\leq R_{00} .
\ee
For unital trace-preserving maps in equal dimensions $d_1=d_2=d$, where ${\bf t}={\bf s}=0$, the above condition simplifies to
\be   \label{I}
\Vert \Lambda \Vert_\infty \leq \frac{1}{d-1}.
\ee
In particular for $d_1=d_2=2$, formula (\ref{I}) reduces to $\Vert \Lambda \Vert_\infty \leq 1$ and defines a necessary and sufficient condition for positivity for trace-preserving unital maps.
\end{COR}
It should be stressed that condition (\ref{main_cond0}) is sufficient but not necessary for a map $\Phi$ to be positive. 

\subsection{Examples of entanglement witnesses and positive maps}

Let us review some well known examples EWs and positive maps in connection to a sufficient condition (\ref{main_cond0}).

\begin{EX} Transposition map $T$ is unital and trace-preserving and hence ${\bf s} = {\bf t} =0$, and $R_{00}=1$.
Taking  $\{\Gamma_\alpha\}$ as generalized Gell-Mann matrices one finds

$$   T(\Gamma_\alpha) = \pm \Gamma_\alpha ,$$
depending on whether $\Gamma_\alpha$ is symmetric or anti-symmetric. Hence  $\Vert \Lambda \Vert_\infty=1$ and the condition (\ref{main_cond0}) is satisfied only for $d_1=d_2=2$.    
\end{EX}

\begin{EX} Reduction map $\Phi : L(H_d) \to L(H_d)$ is defined via 

\be
\Phi(X) = I_d \tr X - X .
\ee
One finds ${\bf s} = {\bf t} =0$ and $R_{00}=d-1$. Now, 

$$   \Phi(\Gamma_k) = - \Gamma_k , \ \ \ k=1,\ldots,d^2-1 ,$$
and hence $\Vert \Lambda \Vert_\infty=1$. Clearly condition (\ref{main_cond0}) is satisfied since

\be
\Vert \Lambda \Vert_\infty = \frac{R_{00}}{d-1} ,
\ee
for all dimensions $d$. Note, that reduction map can be immediately generalized as follows

\be   \label{GG}
\Phi(X) = I_d \tr X - \sum_{k,\ell=1}^{d^2-1} \Lambda_{\ell k} \tr(\Gamma_k X) \Gamma_l   ,
\ee
with arbitrary $\Lambda_{k\ell}$ satisfying $\Vert \Lambda \Vert_\infty\leq 1$. In particular for $\Lambda_{k \ell}$ being an orthogonal matrix. 

\end{EX}

\begin{EX} Consider a map $\Phi : L(H_{d_1}) \to L(H_{d_2})$  defined via

\be   \label{GO}
\Phi(X) = I_{d_2} \tr X - \sum_{\alpha=0}^{d_1^2-1} \sum_{\beta=0}^{d_2^2-1} O_{\beta\alpha} \tr(\Gamma_\alpha X) \Omega_\beta   ,
\ee
where $O$ is an $d_2^2 \times d^2_1$ isometry, i.e. if $d_1 \geq d_2$ one has $OO^{\rm T}  = I_{d_2^2}$.  It provides a generalization of  (\ref{GG}). In the Appendix it is shown that (\ref{GO}) defines a positive map.  
For $\Phi$ defined in (\ref{GO}) one has

$$   R_{00} = \sqrt{d_1 d_2} - O_{00}\ , \ \ s_k = O_{0k} , \ \ t_\ell = O_{\ell 0} \ , \ \ \Lambda_{k \ell} = - O_{k \ell } .  $$
    
\end{EX}

\begin{EX} Any 2-qubit entanglement witness $W$ can be represented as follows

\begin{equation}\label{}
  W = \sum_{\mu,\nu=0}^3 T_{\mu\nu} \sigma_\mu \otimes \sigma_\nu ,
\end{equation}
with $\sigma_\mu \in \{I,\sigma_x,\sigma_y,\sigma_z\}$, and 16 real parameters $T_{\mu\nu}$.
Authors of Ref. \cite{Chiara2020} analyzed a particular scenario when one has an access to the limited resources and consider EW  with diagonal correlation tensor $T_{kl} = c_k \delta_{kl}$ $(k,l=1,2,3)$ only,  that is,

\begin{align} \label{W}
W = \alpha\, I \otimes I +\sum_{k=1}^3 \Big(  t_k \, I \otimes \sigma_k
 + s_k \, \sigma_k\otimes I \Big)  + \sum_{k=1}^3 \lambda_k  \sigma_k\otimes \sigma_k  ,
\end{align}
with real parameters $\alpha,s_k,t_k,\lambda_k$. This reduces the number of independent parameters from 6 elements $T_{kl}=T_{lk}$ to 3 real parameters $\lambda_k$. The above form is justified by the fact that the mean values of single qubit operators $\sigma_{k}\otimes\mathbb{I}$ and
$\mathbb{I}\otimes\sigma_{k}$  for $i=1,2,3$ can be derived by simply ignoring the statistics on one side. There are six 1-parameter families of rank-1 projectors $|\varphi\rangle\langle \varphi|$ of the form (\ref{W}):

\begin{eqnarray}
&\ket{\varphi_{1}}=  a\ket{\phi^{+}}+b\ket{\phi^{-}} ,  \ \ 
 \ket{\varphi_{2}}=  a\ket{\psi^{+}}+b\ket{\psi^{-}} , \ \ 
\ket{\varphi_{3}}=  a\ket{\phi^{+}}+b\ket{\psi^{+}} , \nonumber \\
& \ket{\varphi_{4}}=  a\ket{\phi^{-}}+b\ket{\psi^{-}} ,\ \
\ket{\varphi_{5}}=  a\ket{\phi^{+}}+ib\ket{\psi^{-}} , \ \
\ket{\varphi_{6}}=  a\ket{\phi^{-}}+ib\ket{\psi^{+}} , \nonumber
\end{eqnarray}
where $\ket{\phi^{\pm}}$ and $\ket{\psi^{\pm}}$ are the Bell states, and $a,b \in \mathbb{R}$ such that $a^2+b^2=1$. Hence, there are six 1-parameter families of optimal EWs of the form $W_k := |\varphi_k\rangle\langle \varphi_k|^\Gamma$.
For example one finds for $W_1$: $R_{00} = a^2+b^2=1$, ${\bf s} = {\bf t} = (0,0,2ab)$, and $\lambda_1=-\lambda_2 = a^2 - b^2$, $\lambda_3 =  a^2+b^2=1$. Hence, inequality (\ref{main_cond0}) reduces to  $ 4|ab| + 1 \leq 1$ which implies that either $(a=1,b=0)$ or $(a=0,b=1)$. It shows that only four optimal EWs $|\phi^\pm\rangle\langle \phi^\pm|^\Gamma$ and $|\psi^\pm\rangle\langle \psi^\pm|^\Gamma$ satisfy inequality (\ref{main_cond0}).
\end{EX}

\subsection{A sufficient condition for complete positivity}

As a byproduct of our analysis let us formulate the sufficient condition for complete positivity. 


\begin{proposition}
A positive map (\ref{eq-orthonormal})
is a completely  positive  if the following condition holds:
\be\label{main_cond1}
\Vert{\bf s}\Vert^2+\Vert {\bf t}\Vert^2+\tr(\Lambda\Lambda^T)\leq \frac{R_{00}^2}{d_1d_2-1}.
\ee   
\end{proposition}

\ni {\bf Proof:}
According to a celebrated theorem of Choi \cite{choi1975}, a map is completely positive if its Choi matrix 
\be\label{choi1} C_\Phi=d_1(I_{d_1}\otimes \Phi)(|\phi^+\ra\la \phi^+|)=
\sum_{m,n=1}^{d_1} |m\ra\la n|\otimes \Phi(|m\ra\la n|) ,\ee
 is positive. One finds
\be
C_\Phi= \sum_{\a,\beta} R_{\beta\a}\Gamma_\a^T\otimes \Omega_\beta,
\ee
and hence
\be\label{cphi}
C_{\Phi}=\frac{R_{00}}{\sqrt{d_1 d_2}}I_{d_1}\otimes I_{d_2}+\frac{1}{\sqrt{d_2}}{\bf s}\cdot \Gamma^T\otimes I_{d_2}+\frac{1}{\sqrt{d_1}}I_{d_1}\otimes {\bf t}\cdot \Omega + \sum_{k,i} \Lambda_{ki}\Gamma_i^T\otimes \Omega_k,
\ee
from which we find
\be
\tr(C_{\Phi})=R_{00}\sqrt{d_1d_2},\h \tr(C_{\Phi}^2)=R_{00}^2+\Vert{\bf s}\Vert^2+\Vert {\bf t}\Vert^2 +\tr(\Lambda^T\Lambda).
\ee
Mehta's Lemma \eqref{lemma-mehta} implies  (\ref{main_cond1}) which completes the proof. \hfill $\square$

\begin{EX}
     
 As an example consider the trace-preserving qubit maps $\Phi$ with diagonal elements $\Lambda_{ij} = \delta_{ij} \lambda_i$.  Then the condition for positivity of the map is given from \eqref{main_cond0} to be
\be
\sqrt{t_1^2+t_2^2+t_3^2}+ \lambda_{max}\leq 1,
\ee
and the condition of complete positivity is given from \eqref{main_cond1} to be
\be
\Vert {\bf t}\Vert^2+\sum_i\lambda_i^2\leq \frac{1}{3}.
\ee
We shoud emphasize that since the Mehta's condition is a sufficient conditions, the above inequalities determine only a subset of positive and completely positive maps. For example, if we restrict ourselves to unital maps, then the condition \eqref{main_cond1} is given by $$\lambda_1^2+\lambda_2^2+\lambda_3^2\leq \frac{1}{3},$$ which is a sphere of  radius $r=\frac{1}{\sqrt{3}}$ and is a subset of the Fujiwara-Algoet tetrahedron
$$ |\lambda_1+\lambda_2|\leq 1+\lambda_3\ \ \ \ \   |\lambda_1-\lambda_2|\leq 1-\lambda_3.$$ A comparison of the volumes of these two sets shows that $30$ percents of the completely positive unital maps for qubits are detected by this simple use of Mehta's criteria, figure (\ref{ratio}).
\begin{figure}[!ht]
	\centering
	\includegraphics[width=8.75 cm,height=7.0cm,angle=0]{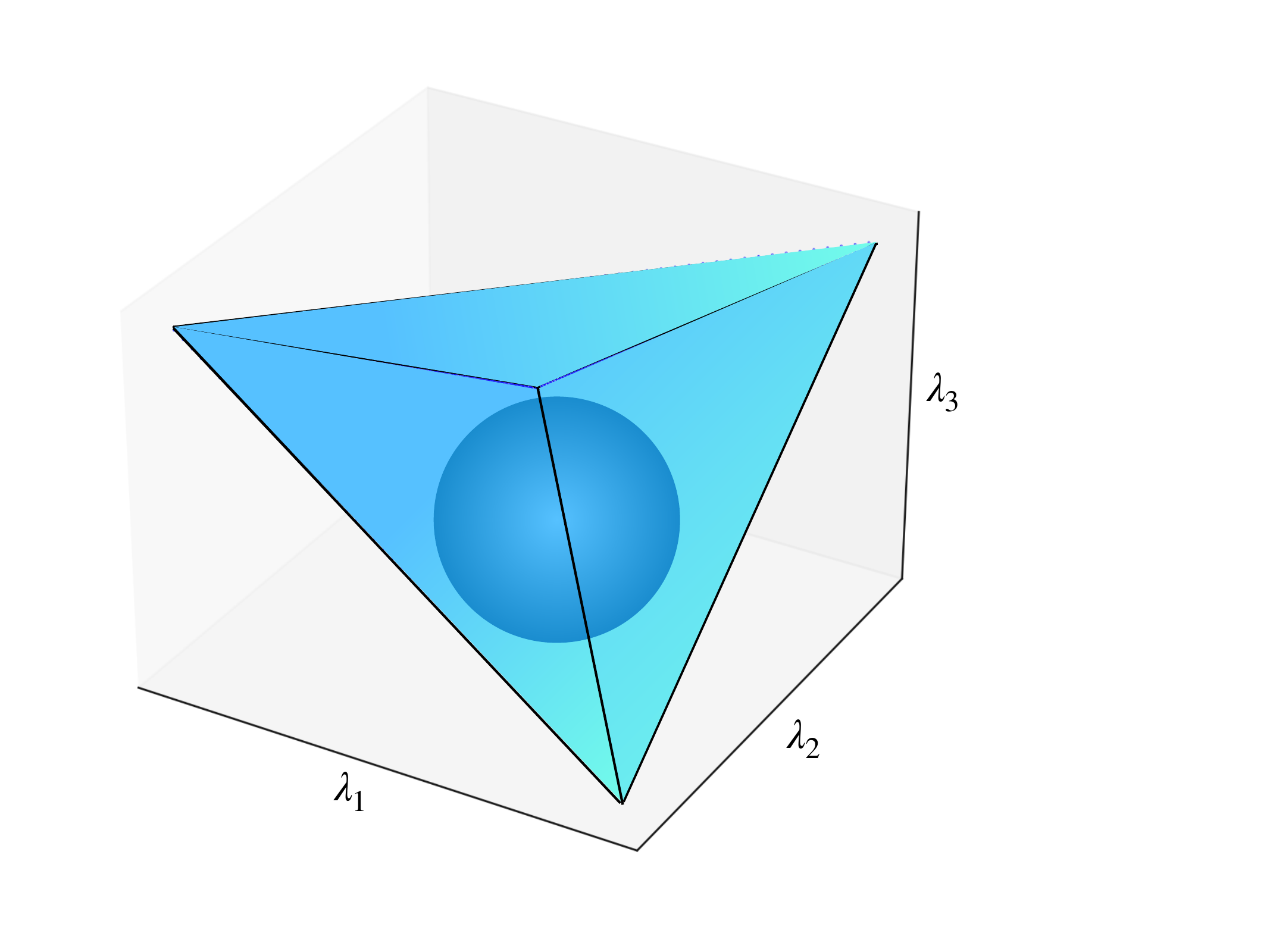}
	\vspace{-.7cm}\caption{ Example of qubit maps, with eigenvalues of affine transformation given by $\lambda_1, \lambda_2$ and $\lambda_3$. Restricting to unital maps, the tethdrahedron shows the complete set of unital maps as derived from the Fujiwara-Algoet condition, the sphere shows those maps which are derived from Mehta's condition. Its volume is approximately 30 percent of the volume of the tethrahedron. 		
	}\label{ratio}
\end{figure}
\end{EX}

\subsection{Mirorred witnesses }

The concept of a mirrored entanglement witness was introduced in Ref. \cite{Bae2020}: given an entanglement witness $W$ one defines a family

\begin{equation}
    W_\mu := \mu I_{d_1} \otimes I_{d_2} - W .
\end{equation}
It is proved \cite{Bae2020} that there exists a minimal $\mu_0$ such that for all $\mu \geq \mu_0$, $W_\mu$ is non-negative on separable states.  If $W_{\rm M}:= W_{\mu_0}$ is an EW one calls $(W,W_{\rm M})$ a mirrored pair of EWs \cite{Bae2020}.

\begin{proposition} If $W$ satisfies (\ref{main_cond0}), then for $\mu \geq \frac{2R_{00}}{\sqrt{d_1 d_2}}$, $W_\mu$ is non-negative on separable states.  
\end{proposition}
Proof: indeed, if $W$ is characterized by $(R_{00},{\bf s},{\bf t},\Lambda)$ an operator $W_\mu$ is characterized by 
$(\mu \sqrt{d_1d_2} - R_{00},-{\bf s},-{\bf t},-\Lambda)$. Hence, if $W$ satisfies (\ref{main_cond0}) and $\mu \sqrt{d_1d_2} - R_{00} \geq R_{00}$, then $W_\mu$ satisfies (\ref{main_cond0}) as well. A condition $\mu \sqrt{d_1d_2} - R_{00} \geq R_{00}$ provides a lower bound $\mu \geq \frac{2R_{00}}{\sqrt{d_1 d_2}} $. \hfill $\Box$

\begin{EX} Consider an EW corresponding to the celebrated Choi map \cite{Bae2020,Anindita2023}

\begin{equation}
    W = \sum_{i=0}^2 \Big( 2|i,i\>\<i,i| + |i,i-1\>\<i,i-1| - 3 |\phi^+\>\<\phi^+| \Big) ,
\end{equation}
where $|\phi^+\> = (|0,0\> + |1,1\> + |2,2\>)/\sqrt{3}$ is a Bell state. One easily finds 
$R_{00}= 2$, ${\bf s}={\bf t}=0$, and $\|\Lambda\|_\infty = 1$, and hence

\begin{equation}
    \|\Lambda\|_\infty = \frac{R_{00}}{2} ,
\end{equation}
that is, the inequality (\ref{main_cond0}) is saturated. One has $ \frac{2R_{00}}{\sqrt{d_1 d_2}} = \frac 43$ and reproduces $\mu_0 = \frac 43$  found in Ref. \cite{Anindita2023}. 
    
\end{EX}

\section{Entanglement Witnesses and realignment-like criterion}   \label{REAL}

Consider a bi-partite state living in $H_{d1} \otimes H_{d_2}$ and represented by

\begin{equation}
    \rho = \sum_{\a,\beta} C_{\beta\a} \Gamma_\a^T \otimes \Omega_\beta .
\end{equation}
Due to the celebrated realignment criterion \cite{Chen1,Chen2,Rudolph} one has

\begin{proposition} If $\rho$ is separable, then

\begin{equation}
    \| C\|_1 \leq 1 . 
\end{equation}

\end{proposition}
Actually, the above criterion was recently generalized as follows \cite{Sarbicki_2020}: let us introduce two square diagonal matrices:

\begin{eqnarray*}
  D_1(x_1) = \mathrm{diag}\{x_1,1,\dots,1\} , \ \
  D_2(x_2) = \mathrm{diag}\{x_2,1,\dots,1\} ,
\end{eqnarray*}
where $D_1(x_1)$ is $d_1^2 \times d_1^2$ and $D_2(x_2)$ is $d_2^2 \times d_2^2$, and the real parameters $x_1,x_2 \geq 0$. One proves \cite{Sarbicki_2020}

\begin{theorem} \label{TH-1} If $\rho$ is separable, then

\begin{equation}\label{xy}
  \| D_1(x_1) C D_2(x_2) \|_{1} \leq \mathcal{N}_{1}(x_1) \mathcal{N}_2(x_2) ,
\end{equation}
where
\begin{equation}\label{NANB}
   \mathcal{N}_1(x_1) = \sqrt{ \frac{d_1 -1 + x_1^2}{d_1}}\ , \ \ \ \ \mathcal{N}_2(x_2)= \sqrt{ \frac{d_2 -1 + x_2^2}{d_2}} ,
\end{equation}
for arbitrary $x_1,x_2 \geq 0$.
\end{theorem}
Note that for $ (x_1, x_2) = (1, 1)$ the inequality (\ref{xy}) reproduces CCNR criterion. Moreover,
for $(x_1, x_2) = (0, 0)$ it reproduces separability criterion derived
by de Vicente \cite{devicente2007separability}. If $d_1 = d_2$, then CCNR criterion is
stronger than the de Vicente criterion \cite{devicente2007separability}. However, for bipartite states
$\rho$ with maximally mixed marginals $\rho_1 = I_{d_1}/d_1$  and $\rho_2 = I_{d_2}/d_2$, the de Vicente criterion is
stronger than CCNR if $d_1 \neq d_2$, and they are equivalent if $d_1 = d_2$ \cite{devicente2007separability}. Here we prove the following result: 

\begin{theorem} The criterion (\ref{xy}) corresponding to $(x_1,x_2)=(0,0)$ is equivalent to the entire class of entanglement witnesses characterized by (\ref{main_cond0}).
\end{theorem}

Proof: consider an EW 
\be\label{eww}
W= \sum_{\a,\beta} R_{\beta\a}\Gamma_\a^T\otimes \Omega_\beta=R_{00}\Gamma_0\otimes \Omega_0 + {\bf s}\cdot \bm \Gamma \otimes \Omega_0+\Gamma_0\otimes {\bf t}\cdot \bm\Omega+ \sum_{k,i} \Lambda_{ki}\Gamma_i^T\otimes \Omega_k ,
\ee
and a bipartite state $\rho\in D(H_{d_1}\otimes H_{d_2})$ be a bi-partite density matrix given by
\be
\rho=\sum_{\a,\beta} C_{\beta\a}\Gamma_\a^T\otimes \Omega_\beta = \frac{1}{\sqrt{d_1d_2}}\Gamma_0\otimes \Omega_0+ {\bf r_1}\cdot \bm \Gamma \otimes \Omega_0+\Gamma_0\otimes {\bf r_2}\cdot \bm\Omega+ \sum_{k,i} Q_{ki}\Gamma_i^T\otimes \Omega_k.
\ee
One finds 
\be
\tr(W\rho)=\frac{R_{00}}{\sqrt{d_1d_2}}+{\bf r_1}\cdot {\bf s}+{\bf r_2}\cdot {\bf t}+\tr(Q\Lambda^T).
\ee
In the above expression, the triple $({\bf r_1}, {\bf r_2}, Q)$ represent the data of the density matrix and the triple $({\bf s}, {\bf t}, \Lambda)$ represent the data of the entanglement witness $W$. Given $\rho$ let us minimize $\tr(W\rho)$ over all EWs from the class (\ref{main_cond0}). one obviously has 
\be
{\bf s}=-x{\bf r_1},\h {\bf t}=-y{\bf r_2},
\ee
where $x$ and $y$ are positive constants. To choose $\Lambda$, we note the singular value decomposition of ${Q}=O{\bf q}O'$, where $O$ and $O'$ are orthogonal matrices and ${\bf q}_{ij} = q_i \delta_{ij}$, where $q_i$ are singular values of $Q$. 
 Let $\Lambda=O'^T\bm \lambda O^T$, where $\bm\lambda_{ij} = - \lambda_{max} \delta_{ij} $. 
One has
\be
\tr(Q\Lambda^T)=-\lambda_{max}\sum_{i}q_i=-\lambda_{max}\Vert Q\Vert_1 .
\ee
Putting everything together, we find
\be
\tr(\rho W)=\frac{R_{00}}{\sqrt{d_1d_2}}-x |{\bf r_1}|^2-y |{\bf r_2}|^2-\lambda_{max}\Vert Q\Vert_1,
\ee
where $\Vert Q\Vert_1$ is the trace norm of $Q$. 
Now we have to minimize this with respect to the data of the entanglement witness $(x,y, \lambda_{max})$ while respecting the condition (\ref{main_cond0}), which now takes the form
\be
y\sqrt{d_2-1}|{\bf r_2}|+x\sqrt{d_1-1}|{\bf r_1}|+\lambda_{max}\sqrt{(d_1-1)(d_2-1)}\leq R_{00}.
\ee
The above condition defines a simplex in the Cartesion space $(x,y,\lambda_{max})$ with vertices $v_0=(0,\ 0,\ 0),$ and 
\be
 v_1= \left(\frac{ R_{00}} { |{\bf r_1}|\sqrt{d_1-1} },\ 0,\ 0\right),\ \ \  v_2=\left(0,\ \frac{ R_{00} } {| {\bf r_2}|\sqrt{d_2-1} },\ 0\right),\ \ \  v_3=\left(0,\ 0,\  \frac{ R_{00} } {\sqrt{(d_1-1)(d_2-1)} }\right).
\ee
 Since $\tr(W\rho)$ is a linear function of $(x,y,\lambda_{max})$, it achieves its minimum on one of the above vertices. One easily finds the value of $\tr(W_i\rho)$ corresponding to the vertex $v_i$
\be
\tr(W_i\rho)=\frac{R_{00}}{\sqrt{d_1d_2}}F_i,\ \ \ i=0,\ 1,\ 2,\ 3,
\ee
where $F_0=1$, and 
\be\label{final-cond}
F_1=1-|{\bf r_1}|\sqrt{\frac{d_1d_2}{d_1-1}},\ \ \ F_2=1-|{\bf r_2}|\sqrt{\frac{d_1d_2}{d_2-1}},\ \ \ \ F_3=1-\Vert Q\Vert_1 \sqrt{\frac{d_1d_2}{(d_1-1)(d_2-1)}}.  
\ee
Positivity of  the reduced density matrices $\rho_1\in D(H_{d_1})$ and $\rho_2\in D(H_{d_2})$ imply that $F_1$ and $F_2$ are always non-negative. The condition $F_3 \geq 0$ is equivalent to

\begin{equation}  \label{Q1}
    \|Q\|_1 \leq \sqrt{\frac{(d_1-1)(d_2-1)}{d_1d_2}}  ,
\end{equation}
which reproduces (\ref{xy}) for $(x_1,x_2)=(0,0)$. \hfill $\Box$


\begin{EX}
 
In Ref. \cite{Bae2022} the following family of PPT entangled states in $C^4\otimes C^4$ is introduced
\be\label{rho-4}
\rho = \frac{1}{N}\left(
\begin{array}{cccc|cccc|cccc|cccc}
	1 & . & . & . & . & 1 & . & . & . & . & 1 & . & . & . & . & 1 \\
	. & \frac{1}{a} & . & . & . & . & . & . & . & . & . & . & . & . & 1 & . \\
	. & . & 1 & . & . & . & . & . & 1 & . & . & . & . & . & . & . \\
	. & . & . & a & . & . & 1 & . & . & . & . & . & . & . & . & . \\ \hline
	. & . & . & . & a & . & . & . & . & . & . & 1 & . & . & . & . \\
	1 & . & . & . & . & 1 & . & . & . & . & 1 & . & . & . & . & 1 \\
	. & . & . & 1 & . & . & \frac{1}{a} & . & . & . & . & . & . & . & . & . \\
	. & . & . & . & . & . & . & 1 & . & . & . & . & . & 1 & . & . \\ \hline
	. & . & 1 & . & . & . & . & . & 1 & . & . & . & . & . & . & . \\
	. & . & . & . & . & . & . & . & . & a & . & . & 1 & . & . & . \\
	1 & . & . & . & . & 1 & . & . & . & . & 1 & . & . & . & . & 1 \\
	. & . & . & . & 1 & . & . & . & . & . & . & \frac{1}{a} & . & . & . & . \\ \hline
	. & . & . & . & . & . & . & . & . & 1 & . & . & \frac{1}{a} & . & . & . \\
	. & . & . & . & . & . & . & 1 & . & . & . & . & . & 1 & . & . \\
	. & 1 & . & . & . & . & . & . & . & . & . & . & . & . & a & . \\
	1 & . & . & . & . & 1 & . & . & . & . & 1 & . & . & . & . & 1 \\
\end{array}\right) ,
\ee
where dots represent zeros, and $N = 4 a+\frac{4}{a}+8$ stands for the normalization factor, 
and it is shown that their entanglement can be detected by an entanglement witness, called $W_{ext}$, which shows that these states are entangled in the interval $0<a<1$. We now show that the interval of entanglement of these states is in fact much larger and covers the whole positive line $0<a<\infty$, except the point $a=1$.
To show this, the $Q$ matrix corresponding to \eqref{rho-4} is calculated, which turns out to be
\be\nonumber
Q=\frac{1}{(a+1)^2} \big(A\oplus A'\oplus B\big),
\ee
where
\be\nonumber
A = \left(\begin{array}{cccccc}
	\frac{a}{4 } & . & . & . & . & \frac{a}{4 }  \\
	. & \frac{a}{2 } & . & . & . & .  \\
	. & . & \frac{a}{4 } & \frac{a}{4 } & . & .  \\
	. & . & \frac{a}{4 } & \frac{a}{4 } & . & .  \\
	. & . & . & . & \frac{a}{2 } & . \\
	\frac{a}{4 } & . & . & . & . & \frac{a}{4 }
	\end{array}
\right),\ \ \ 
A' = \left(\begin{array}{cccccc}
 -\frac{a}{4 } & . & . & . & . & \frac{a}{4 }  \\
 . & . & . & . & . & . \\
	 . & . & -\frac{a}{4 } & -\frac{a}{4 } & . & .  \\
	 . & . & -\frac{a}{4 } & -\frac{a}{4 } & . & . \\
 . & . & . & . & . & . \\
 \frac{a}{4 } & . & . & . & . & -\frac{a}{4 }
\end{array}
\right), \ \ 
%
B = \left(\begin{array}{ccc}
	-\frac{(a-1)^2}{8 } & -\frac{a^2+2 a-3}{8 \sqrt{3} } & -\frac{(a-1) a}{2 \sqrt{6} } \\
	\frac{3 a^2-2 a-1}{8 \sqrt{3} } & -\frac{(a-1)^2}{24 } & -\frac{a^2+a-2}{6 \sqrt{2} } \\
	\frac{a-1}{2 \sqrt{6} } & \frac{2 a^2-a-1}{6 \sqrt{2} } & -\frac{(a-1)^2}{12 }
\end{array}
\right),
\ee
\ni with singular values given as
\be\nonumber
\Big\{  \frac{\left| a-1\right| }{4 \left| a+1\right| } \  (\times 2),\  \frac{\left| a\right| }{2 (a+1)^2} \  (\times 6), \  \frac{(a-1)^2}{4 (a+1)^2}\  \Big\}.
\ee
This leads to
\be\nonumber
 \Vert Q \Vert_{1} = \frac{1}{4}  \frac{(a-1)^2}{(a+1)^2} +3 \frac{|a|}{(a+1)^2} +\frac{1}{2}  \left|\frac{a-1}{ a+1 }\right| .
\ee
Figure \ref{fig-alpha.gamma} shows that the value of $\gamma= \Vert Q \Vert_{1}$ is always larger than $\sqrt{\frac{(d_1-1)(d_2-1)}{d_1d_2}}=\frac{3}{4}$ except for $a=1$.
\begin{figure}[!ht]
	\centering
	\includegraphics[width=10cm,height=10cm,angle=0]{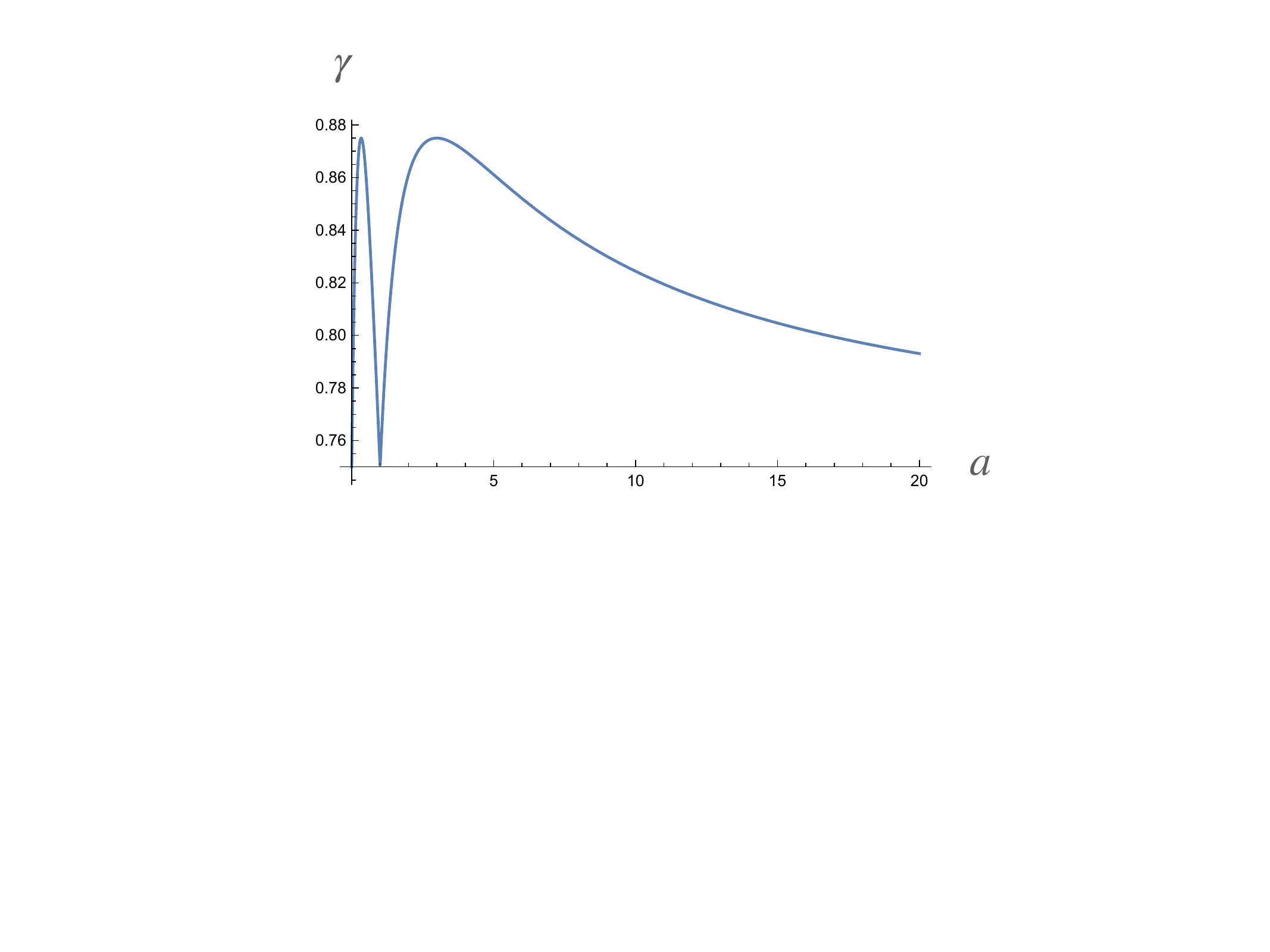}
	\vspace{-4.5cm}
    \caption{
	A state \eqref{rho-4} is entangled for $0 < a < \infty;  \ a\neq 1$.
	}\label{fig-alpha.gamma}
\end{figure}
Note that by $a<\infty$, we are emphasizing that $a$ should be a finite value. In fact, in the limit $a\to \infty$, the state approaches to
\be\nonumber
\rho_{\infty}=\frac{1}{4}\Big(|03\ra\la 03|+|10\ra\la 10|+|11\ra\la 11|+|22\ra\la 22| \Big),
\ee
which is obviously separable. Incidentally, this is the limiting case where $\gamma  $ approaches the value $\frac{3}{4}$.

\end{EX}

\begin{EX}
   
In Ref. \cite{Terhal2002} Terhal introduced a family of UPB for $3\otimes n$ systems. Here we take $n=4$ and show that our criterion (\ref{Q1}) can detect the corresponding PPT entangled state in $3\otimes 4 $. 
The UPB states for the $3\times 4$ system are:

\begin{eqnarray*}
&\ket{\psi_0} = \frac{1}{\sqrt{2}}|0\rangle \otimes (|1\rangle-|3\rangle),\ \ 
\ket{\psi_1} = \frac{1}{\sqrt{2}} |1\rangle \otimes (|2\rangle-|3\rangle),\ \ 
\ket{\psi_2} = \frac{1}{\sqrt{2}}|2\rangle \otimes (|0\rangle-|3\rangle),\ \ 
\ket{\psi_3} = \frac{1}{\sqrt{2}}(\ket{0}-\ket{1}) \otimes |0\rangle, \\
& \ket{\psi_4} = \frac{1}{\sqrt{2}}(\ket{1}-\ket{2})\otimes |1\rangle,\ \ 
\ket{\psi_5} = \frac{1}{\sqrt{2}}(\ket{2}-\ket{0})\otimes |2\rangle, \ \ 
\ket{\psi_6} =\frac{1}{\sqrt{12}}\sum_{i=0}^2\sum_{j=0}^{3}|i\rangle\otimes |j\rangle .
\end{eqnarray*}
The PPT entangled state $\rho= \frac 16(I-\sum_{i=0}^6|\psi_i\ra\la \psi_i|)$ turns out to be
\ba\label{rho-6}
 \rho = \frac{1}{60}\left(
\begin{array}{ccc|ccc|ccc|ccc}
	5 & -1 & -1 & -1 & 5 & -1 & -1 & -1 & -1 & -1 & -1 & -1 \\
	-1 & 5 & -1 & 5 & -1 & -1 & -1 & -1 & -1 & -1 & -1 & -1 \\
	-1 & -1 & 5 & -1 & -1 & -1 & -1 & -1 & -1 & -1 & 5 & -1 \\ \hline
	-1 & 5 & -1 & 5 & -1 & -1 & -1 & -1 & -1 & -1 & -1 & -1 \\
	5 & -1 & -1 & -1 & 5 & -1 & -1 & -1 & -1 & -1 & -1 & -1 \\
	-1 & -1 & -1 & -1 & -1 & 5 & -1 & -1 & -1 & 5 & -1 & -1 \\ \hline
	-1 & -1 & -1 & -1 & -1 & -1 & 5 & 5 & -1 & -1 & -1 & -1 \\
	-1 & -1 & -1 & -1 & -1 & -1 & 5 & 5 & -1 & -1 & -1 & -1 \\
	-1 & -1 & -1 & -1 & -1 & -1 & -1 & -1 & 5 & -1 & -1 & 5 \\ \hline
	-1 & -1 & -1 & -1 & -1 & 5 & -1 & -1 & -1 & 5 & -1 & -1 \\
	-1 & -1 & 5 & -1 & -1 & -1 & -1 & -1 & -1 & -1 & 5 & -1 \\
	-1 & -1 & -1 & -1 & -1 & -1 & -1 & -1 & 5 & -1 & -1 & 5 \\
\end{array}
\right).
\ea
The corresponding $Q$ matrix is given by
\be\nonumber
Q=\frac{1}{30}\left(
\begin{array}{c|c|c}
- \mathbb{J}_{3\times 6} & 0_{3\times 6} & A \\ \hline
0_{3\times 6} & 0_{3\times 6} & 0_{3\times 3} \\ \hline
B & 0_{2\times 6} & 0_{2\times 3}
\end{array}
\right) ,
\ee
where
\be\nonumber
A =\left(
\begin{array}{ccc}
 3 & \sqrt{3} & \sqrt{\frac{3}{2}} \\
0 & -2\sqrt{3} & \sqrt{\frac{3}{2}} \\
-3& -\sqrt{3} & \sqrt{\frac{3}{2} }
\end{array}
\right) , \ \ \
B=\left(
\begin{array}{cccccc}
	0 & 0 & 0 & 0 & 3 & -3 \\
	0 & 0 & 0 & -2\sqrt{3} & \sqrt{3} &  \sqrt{3} \\
\end{array}
\right) ,
\ee
and the $3 \times 6$ matrix $\mathbb{J}_{ij}=1$. 
The singular values of $Q$ are found to be
\be\nonumber
 \left\{\frac{1}{2 \sqrt{10}},\frac{1}{5 \sqrt{2}},\frac{1}{5 \sqrt{2}},\frac{1}{5 \sqrt{2}},\frac{1}{5 \sqrt{2}}\right\} ,
\ee
 and we find
 \be\nonumber
  \Vert Q \Vert_{1} =\frac{1}{2\sqrt{10}}+\frac{4}{5\sqrt{2}}= \frac{8+\sqrt{5}}{10\sqrt{2}},
 \ee
 which is greater than $ \sqrt{\frac{(d_1-1)(d_2-1)}{d_1d_2}}=\frac{1}{\sqrt{2}}$.
 This shows that the PPT state \eqref{rho-6} is entangled.

\end{EX}

\section{Conclusion}  \label{CON}

Mehta's lemma provides a simple sufficient condition for a matrix to be positive. We have used this lemma to construct a large family of positive maps and characterized them in terms of their corresponding affine transformations, i.e. transformation of the so-called Bloch vector ${\bf r}\lo \Lambda {\bf r}+ x_0{\bf t}$. Using  one-to-one correspondence between entanglement witnesses and the Choi matrices of positive maps, a multi-parameter family of Entanglement Witnesses for bipartite systems with different dimensions was constructed.  Interestingly, it is shown that this class of witnesses is equivalent to a single realignment-like criterion already analyzed in \cite{devicente2007separability}.
The power of this criterion is illustrated by two examples of PPT entangled states in $4 \otimes4$ and $3 \otimes 4$ systems.


\section*{Acknowledgement} This research was supported in part by Iran National Science Foundation, under Grant No.4022322. The authors wish to thank S. Roofeh, A. Farmanian,  and A. Tangestani,  for many inspiring discussions during this project. DC was partially supported by Polish National Science Center project No. 2018/30/A/ST2/00837.

\bibliographystyle{unsrt}

\bibliography{Refs}

\appendix

\section{Proof of positivity of (\ref{GO})}

One finds for normalized $\phi \in H_{d_1}$ and $\psi \in H_{d_2}$:  

$$   \la \psi |\Phi(|\phi\ra \la \phi|)|\psi\ra = 1 - \sum_{\beta=0}^{d_2^2-1} \la \phi| \tilde{\Gamma}_\beta | \phi\ra  \la \psi|\Omega_\beta|\psi\ra   ,  $$
where 
$$ \tilde{\Gamma}_\beta =  \sum_{\alpha=0}^{d_1^2-1} O_{\beta\alpha} \Gamma_\alpha .$$
Now, using Cauchy-Schwarz inequality one obtains

$$\sum_{\beta=0}^{d_2^2-1} \la \phi \tilde{\Gamma}_\beta | \phi\ra  \la \psi\Omega_\beta|\psi\ra  \leq   
\sqrt{ \sum_{\beta=0}^{d_2^2-1} |\la \phi| \tilde{\Gamma}_\beta | \phi\ra|^2  }  \sqrt{ \sum_{\beta=0}^{d_2^2-1} | \la \psi|\Omega_\beta|\psi\ra|^2 } . $$
Now, recalling

$$ \sum_{\alpha=0}^{d_1^2-1}  \Gamma^{\rm T}_\alpha \otimes \Gamma_\alpha  =  d_1 P^+_{d_1} , \ \ \ 
 \sum_{\beta=0}^{d_2^2-1}  \Omega^{\rm T}_\beta \otimes \Omega_\beta  =  d_2 P^+_{d_2} ,$$
where $P^+_{d_1}$ and $P^+_{d_2}$ stand for maximally entangled state in $H_{d_1} \otimes H_{d_1}$ and $H_{d_1} \otimes H_{d_2}$, respectively, one finds

$$ \sum_{\beta=0}^{d_2^2-1} |\la \phi| \tilde{\Gamma}_\beta | \phi\ra|^2  \leq 1 ,  \ \ \ 
  \sum_{\beta=0}^{d_2^2-1} | \la \psi|\Omega_\beta|\psi\ra|^2  = 1 , $$
and hence 

$$   \la \psi |\Phi(|\phi\ra \la \phi|)|\psi\ra \geq 0 , $$
which proves positivity of the map $\Phi$. 

\section{An alternative construction of Entanglement Witnesses}\label{sec-alternative}

In this appendix, without resorting to positive maps, we construct an Entanglement Witness $W$ in a direct manner. We use an alternative definition of an Entanglement Witness as an operator which is block-positive but not entirely positive. A block-positive operator $W$ is one which satisfies
$\la \a\otimes \la \beta |W|\a\ra\otimes \beta\ra$ for all product states $|\a\ra\otimes |\beta\ra$, but is not entirely positive and has negative eigenvalues on the full Hilbert space. \\

\ni {\bf Sufficient condition for block-positivity:} To this end, we first apply Mehta's lemma to derive sufficient conditions for block-positivity of $W$. We will later show that for detecting the entanglement witness of a given state, unital entanglement witnesses (i.e. those which correspond to unital postive map) act better than any other witnesses. So from (\ref{eww})
we take a unital $W$ to be
\be\nonumber
W=R_{00}\Gamma_0\otimes \Omega_0+{\bf s}\cdot \bm\Gamma\otimes \Omega_0+ \Gamma_0\otimes {\bf t}\cdot\bm\Omega+\Lambda_{\beta, \alpha}\Gamma_\alpha^T\otimes \Omega_\beta.
\ee
Block-positivity means that for any vector $|v\ra\in H_{d_2}$, the following matrix should be positive
\be\nonumber
B \equiv \la v|W|v\ra=(\frac{R_{00}}{\sqrt{d_2}}+\la v|{\bf t}\cdot \bm\Omega|v\ra)\Gamma_0+\frac{1}{\sqrt{d_2}}{\bf s}\cdot \bm\Gamma+\Lambda_{\beta,\alpha}\Gamma_\a^T\la v|\Omega_\beta|v\ra.
\ee
Using the notation $q_\beta:=\la v|\Omega_\beta|v\ra$, and ${\bf q}= (q_1,q_2,\cdots q_{d_2})$ we have
\be
\tr(B)=\sqrt{d_1}(\frac{R_{00}}{\sqrt{d_2}}+{\bf t}\cdot {\bf q}),
\ee
and
\be\nonumber
\tr(B^2)= (\frac{R_{00}}{\sqrt{d_2}}+{\bf t}\cdot {\bf q})^2+\Vert \frac{{\bf s}}{\sqrt{d_2}}+\Lambda {\bf q}\Vert^2
\ee
where we have used the relations $\tr(\Gamma_\a)=0,\ \ \ \tr(\Gamma_\a\Gamma_\beta)=\delta_{\a\beta}$. We now resort to the Mehta's Lemma and demand that the following condition $\tr(B^2)\leq \frac{\tr(B)^2}{d_1-1}$ holds for all the vectors $|v\ra$.
With a little rearrangement, this leads to 
\be\nonumber
\Vert \frac{{\bf s}}{\sqrt{d_2}}+\Lambda {\bf q}\Vert^2\leq \frac{(\frac{R_{00}}{\sqrt{d_2}}+{\bf t}\cdot {\bf q})^2}{d_1-1}
\forall\ {\bf q}.
\ee
or 
\be
\Vert \frac{{\bf s}}{\sqrt{d_2}}+\Lambda {\bf q}\Vert\leq \frac{(\frac{R_{00}}{\sqrt{d_2}}+{\bf t}\cdot {\bf q})}{\sqrt{d_1-1}}
\h \forall\ {\bf q}.
\ee
To strengthen this sufficient condition, we replace the left hand side of this inequality with a larger term and the right hand side with a smaller one to arrive at 
\be
\frac{\Vert {\bf s}\Vert}{\sqrt{d_2}}+ \Vert \Lambda\Vert_\infty \Vert {\bf q}\Vert \leq  \frac{ \frac{R_{00}}{\sqrt{d_2}} -\Vert {\bf t}\Vert \Vert {\bf q}\Vert } {\sqrt{d_1-1}}.
\ee
Inserting the value $\Vert {\bf q}\Vert=\sqrt{1-\frac{1}{d_2}}$ and simplifying, we arrive at the condition (\ref{main_cond0}), namely
\be\label{main_cond0-a}
\sqrt{d_2-1}\Vert {\bf t}\Vert+\sqrt{d_1-1}\Vert {\bf s}\Vert + \sqrt{(d_1-1)(d_2-1)}\Vert \Lambda\Vert_{\infty}
\leq R_{00} , 
\ee
Just for completeness, we remind the reader of how $\Vert {\bf q}\Vert$ is calculated. Using the summation convention for repeated indices, we first note that 
$$
{\bf q}.{\bf q} = \la v|\Omega_\mu|v\ra\la v|\Omega_\mu|v\ra= \la v,v|\Omega_\mu\otimes \Omega_\mu|v,v\ra.
$$
Using the identity
\be\label{eq-permutation}
\frac{1}{d}I+\Omega_\mu\otimes \Omega_\mu=P ,
\ee
where $P$ is the permutation operator $P|\mu,\nu\ra=|\nu,\mu\ra$, we find that
\be\nonumber
{\bf q}.{\bf q} = \la v, v|P-\frac{1}{d_2}I|v,v\ra= 1-\frac{1}{d_2}.
\ee

\end{document}